\documentclass[12pt, a4paper]{article}
\usepackage{a4wide}
\begin{document}
\title{Deutsch-Jozsa Algorithm Revisited in the Domain of Cryptographically
Significant Boolean Functions}
\author{
Subhamoy Maitra and Partha Mukhopadhyay\\
Applied Statistics Unit, Indian Statistical Institute,\\
203 B. T. Road, Calcutta, Pin 700 108, INDIA\\ 
Communicating E-mail: subho@isical.ac.in
}
\date{}
\maketitle
\newcommand{\qed}{\hfill \rule{2mm}{2mm}}
\newcommand{\pf}{{\bf Proof : }}
\newtheorem{problem}{Problem}
\newtheorem{algorithm}{Algorithm}
\newtheorem{definition}{Definition}
\newtheorem{construction}{Construction}
\newtheorem{theorem}{Theorem}
\newtheorem{question}{Question}
\newtheorem{lemma}{Lemma}
\newtheorem{proposition}{Proposition}
\newtheorem{remark}{Remark}
\newtheorem{corollary}{Corollary}
\newtheorem{example}{Example}
\newcommand{\binom}[2] {\mbox{$\left( { #1 \atop #2 } \right)$}}
\newcommand{\comb}[2] {\mbox{$\left( { #1 \atop #2 } \right)$}}

\begin{abstract}
Boolean functions are important building blocks in cryptography for their
wide application in both stream and block cipher systems. For cryptanalysis
of such systems one tries to find out linear functions that are
correlated to the Boolean functions used in the crypto system. 
Let $f$ be an $n$-variable Boolean function and its Walsh spectra 
is denoted by $W_f(\omega)$ at the point $\omega \in \{0, 1\}^n$.
The Boolean function is available in the form of an oracle. We like to find an 
$\omega$ such that $W_f(\omega) \neq 0$ as this will provide one of the 
linear functions which are correlated to $f$. We show that the
quantum algorithm proposed by Deutsch and Jozsa (1992) solves the above
mentioned problem in constant time. However, the best known classical 
algorithm to solve this problem requires exponential time in $n$. We also
analyse certain classes of cryptographically significant Boolean functions 
and highlight how the basic Deutsch-Jozsa algorithm performs on them.
\end{abstract}


\section{Introduction}
\label{intro}
Many of the symmetric (private) key crypto systems use nonlinear Boolean 
functions in the design process. Nonlinearity is an important property of 
Boolean functions to resist
the linear cryptanalysis~\cite{MM94} on block cipher systems like DES.
Apart from nonlinearity, the Boolean functions should 
also possess other cryptographic properties. In the nonlinear combiner 
model of stream cipher systems, correlation immunity is an important 
cryptographic property for a Boolean function to be used in the 
schemes~\cite{SG84,SG85}. Both the nonlinearity and the correlation immunity 
can be described in terms of Walsh spectra of the Boolean function (see
Subsection~\ref{subbool} for exact details). Construction of 
highly nonlinear and correlation immune Boolean functions are available in 
literature (see~\cite{SG84,SM00A,SM00B,SM00E,CC02} and the references in these
papers). Even if a Boolean function is highly nonlinear and 
correlation immune of certain order, due to the Parseval's 
relation~\cite{CD91}, there always exist linear functions which are correlated
to the Boolean function in use. In the design, it is always attempted to reduce
the correlation, which is the the job of the cryptographer. On the other hand,
the cryptanalyst tries to exploit the correlation to mount the 
attack (see~\cite{AM00,TH00B} and the references in these papers for more 
details). To device such an attack, one needs a linear function which is 
correlated to the Boolean function. Given an $n$-variable Boolean
function $f$, this requires the Walsh spectra of the Boolean function 
and the Fast Walsh Transform algorithm requires $O(n2^n)$ time when the 
truth table of the Boolean function is available. If the Boolean function is 
available in the form of an oracle (black box), then $2^n$ steps are required
to get the truth table and then only the Fast Walsh Transform can be applied.
This is the best known classical algorithm known in this area.
On the other hand we identify that the well known Deutsch-Jozsa 
algorithm~\cite{qDJ92} can solve this problem in constant time under the
quantum computational framework. It has been commented~\cite[Page 36]{qNC02} 
that the Deutsch-Jozsa algorithm 
has not much application in practical sense. This is the first time we show 
how this algorithm can be used to solve a problem which naturally comes from
cryptographic domain.

Now we like to point out the importance of the problem from the quantum 
complexity theoretic viewpoint. For detailed discussion on complexity 
classes and their hierarchies see~\cite{qNC02,DK00}.
The Deutsch-Jozsa problem~\cite{qDJ92} (distinguishing between
balanced and constant Boolean functions) presents relativized 
separation of P and EQP, but not of BPP and BQP. In~\cite{BV93}, Bernstein
and Vazirani presented relativized separation between BPP and BQP using
recursive Fourier sampling. Though the problem is important from complexity
theoretic point of view, it has been commented to be artificial~\cite{SA02}.
Bernstein and Vazirani~\cite{BV97} have further shown the relativized 
separation of NP and even MA from BQP and conjectured that recursive
Fourier sampling is not in PH (related discussion is also available 
in~\cite{SA02}). Green and Pruim~\cite{GP01} presented relativized separation 
between BQP and ${\mbox P}^{\mbox{NP}}$ using a nice technique based on
Grover's algorithm~\cite{qGR96}. Aaronson has commented in~\cite{SA02}
that it may need a completely different problem than recursive Fourier
sampling to provide a relativized separation between BQP and PH. 
The problems we mention here (specifically see Problem~\ref{prob5} in 
Section~\ref{corr}) may be a good candidate in this direction.

\subsection{Preliminaries: Boolean Functions}
\label{subbool}
A Boolean function on $n$ variables may be viewed as a mapping
from $\{0, 1\}^n$ into $\{0, 1\}$. The set of all $n$-variable Boolean 
functions is denoted by $\Omega_n$.

A Boolean function $f(x_1, \ldots, x_n)$ is also interpreted as the output
column of its {\em truth table} $f$, i.e., a binary string of length $2^n$,
$$f = [f(0, 0, \cdots, 0), f(1, 0, \cdots, 0), f(0, 1, \cdots, 0), 
\ldots, f(1, 1, \cdots, 1)].$$ If a Boolean function is presented
as an oracle (a black box), then one can only present an $n$-bit input and 
get the $1$-bit output corresponding to that. Thus, to get the truth table,
one needs to query the oracle $2^n$ times in a classical computational model.

The {\em Hamming distance} between $S_1, S_2$
is denoted by $d(S_1, S_2)$, i.e., $d(S_1, S_2) = \#(S_1 \neq S_2).$
Also the {\em Hamming weight} or simply the weight of a binary string
$S$ is the number of ones in $S$. This is denoted by $wt(S)$.
An $n$-variable function $f$ is said to be {\em balanced} if its output
column in the truth table contains equal number of 0's and 1's
(i.e., $wt(f) = 2^{n-1}$).

Let us denote addition operator over $GF(2)$ by $\oplus$. 
An $n$-variable Boolean function $f(x_1, \ldots, x_n)$ can be considered
to be a multivariate polynomial over $GF(2)$. This polynomial can be expressed
as a sum of products representation
of all distinct $k$-th order products $(0 \leq k \leq n)$ of the
variables. More precisely, $f(x_1, \ldots, x_n)$ can be written as
$$a_0 \oplus \bigoplus_{1 \leq i \leq n} a_i x_i \oplus
\bigoplus_{1 \leq i < j \leq n} a_{ij} x_i x_j \oplus \cdots \oplus
a_{12\ldots n} x_1 x_2 \ldots x_n,$$ where the coefficients
$a_0, a_{ij}, \ldots, a_{12\ldots n} \in \{ 0, 1\}$.
This representation of $f$ is called the {\em algebraic normal form} (ANF) of
$f$. The number of variables in the highest order product term with
nonzero coefficient is called the {\em algebraic degree}, or simply the
degree of $f$ and denoted by $deg(f)$. 

Functions of degree at most one are called {\em affine} functions. An affine
function with constant term equal to zero is called a {\em linear} function.
The set of all $n$-variable affine (respectively linear) functions is denoted
by $A(n)$ (respectively $L(n)$). The nonlinearity of an $n$-variable function
$f$ is $$nl(f) = min_{g \in A(n)} (d(f, g)),$$ i.e., 
the distance from the set of all $n$-variable affine functions.

Let ${x} = (x_1, \ldots, x_n)$ and
${\omega} = (\omega_1, \ldots, \omega_n)$ both belong
to $\{0, 1\}^n$ and the inner product $${x} \cdot {\omega} =
x_1\omega_1 \oplus \cdots \oplus x_n\omega_n.$$
Let $f({x})$ be a Boolean function on $n$ variables.
Then the {\em Walsh transform} of $f({x})$ is a real valued function
over $\{0, 1\}^n$ which is defined as $$W_f({\omega}) = \sum_{{x} 
\in \{0, 1\}^n} (-1)^{f({x}) \oplus {x} \cdot {\omega}}.$$

Given a Boolean function $f$, $W_f(\omega) = \#(f = l) - \#(f \neq l)$,
where $l = \omega \cdot x$ is a linear function. If $W_f(\omega) = 0$,
then there is no correlation between $f$ and $l$. However, if 
$W_f(\omega) > 0$, then there is correlation between $f, l$ as 
$\#(f = l) > \#(f \neq l)$. Similarly, if 
$W_f(\omega) < 0$, then there is correlation between $f, 1 \oplus l$ as 
$\#(f = l) < \#(f \neq l)$, which gives
$\#(f = 1 \oplus l) > \#(f \neq 1 \oplus l)$. This correlation between 
the Boolean function $f$ and the linear function $l$ (or the affine function
$1 \oplus l$) is exploited for cryptanalytic attacks~\cite{AM00,TH00B}.
Thus, given a Boolean function $f$, it is important to find out some
$\omega$ such that $W_f(\omega) \neq 0$.

It should be noted that getting the Walsh spectra is not an easy problem in 
general. See Algorithm~\ref{algo0} in this Section and Proposition~\ref{proph} 
in Section~\ref{corr} later for further discussion.

In terms of Walsh spectra, the nonlinearity of $f$ is given by
$$nl(f)=2^{n-1} - \frac{1}{2}\max_{\omega \in \{0, 1\}^n}|W_f(\omega)|.$$

One important identity related to the Walsh spectra of any $n$-variable 
Boolean function $f$ is the Parseval's identity~\cite{CD91} which gives 
$$\sum_{\omega \in \{0, 1\}^n} W_f^2(\omega) = 2^{2n}.$$ It is clear that
the maximum nonlinearity is achieved when the maximum absolute
value of the Walsh spectra is minimized. For $n$ even, this happens when 
$W_f(\omega) = \pm 2^{\frac{n}{2}}$, for each $\omega \in \{0, 1\}^n$.
These functions, having nonlinearity $2^{n-1} - 2^{\frac{n}{2}-1}$, 
are well known as bent functions in 
literature~\cite{RO76,CC03}. For $n$ odd, $\frac{n}{2}$ is not 
an integer and hence the situation becomes more complicated. For $n \leq 7$,
it is known that the maximum possible nonlinearity can be 
$2^{n-1} - 2^{\frac{n-1}{2}}$~\cite{JJ83}. It has been shown 
in~\cite{PW83} that one can achieve nonlinearity strictly greater than
$2^{n-1} - 2^{\frac{n-1}{2}}$ for $n \geq 15$. 

In~\cite{XZ88}, an important characterization of resilient (balanced and 
correlation immune) functions has been presented, which we use as the 
definition here. A function $f(x_1, \ldots, x_n)$ is $m$-resilient
iff its Walsh transform satisfies $$W_f({\omega}) = 0,
\mbox{ for } 0 \leq wt({\omega}) \leq m.$$
As the notation used in~\cite{SM00A,SM00B}, by an $(n, m, d, \sigma)$ 
function we denote an $n$-variable, $m$-resilient
function with degree $d$ and nonlinearity $\sigma$. 
For recent results on such functions see~\cite{SM00A,SM00B,CC02} and the
references in these papers. 

Now let us present the best known classical algorithm for calculating the 
Walsh spectra of a Boolean function. If the function is given as a black
box, then one needs to get the truth table first, which requires $2^n$
many query to the oracle.

\begin{algorithm} \
\label{algo0}
\begin{center}
\begin{tabular}{|ll|}
\hline
Input: & \ \\
(i) & A Boolean function $f$ on $n$ variables is \\
\   & available in the form of an oracle (black box);\\
\hline
1. & Oracle $f$ is queried $2^n$ many times to get the truth\\
\   & table as an integer array $f[0, \ldots, 2^n-1]$ of $0, 1$;\\
2. & for $(i = 0; i < 2^n; i = i+1)$ $f[i] = (-1)^{f[i]}$; \\
3. & for $(i = 0; i < n; i = i+1)$ \{ \\
3a. & ~~~~~~for $(k = 0; k < 2^n; k = k+2^{i+1})$  \{ \\
3a(i). & ~~~~~~~~~~~~for $(j = k; j < k+2^i; j = j + 1)$  \{ \\
3a(i)A. & ~~~~~~~~~~~~~~~~~~$a = f[j] + f[j+2^i]$; \\
3a(i)B. & ~~~~~~~~~~~~~~~~~~$b = f[j] - f[j+2^i]$; \\
3a(i)C. & ~~~~~~~~~~~~~~~~~~$f[j] = a$; \\
3a(i)D. & ~~~~~~~~~~~~~~~~~~$f[j+2^i] = b$; \\
3a(ii). & ~~~~~~~~~~~~\} \\
3b. & ~~~~~~\} \\
4. & \} \\
\hline
\end{tabular}
\end{center}
\end{algorithm}

In the following we present an example how the Algorithm~\ref{algo0}
runs. Note that the function used is a 3-variable one, and $i$ varies
from $0$ to $2$, i.e., $n = 3$ steps. The inner steps (using $k, j$)
runs $2^3 = 8$ many times.
\begin{center}
\begin{tabular}{|c|c|c||c|||c|c|c|c|}
\hline
$x_3$ & $x_2$ & $x_1$ & $f$ & $(-1)^f$ & $i = 0$ & $i = 1$ & $i = 2$\\
\hline
0&0&0&1&-1&-2&-2&0\\
0&0&1&1&-1&0&2&0\\
0&1&0&0&1&0&-2&-4\\
0&1&1&1&-1&2&-2&4\\
1&0&0&1&-1&0&2&-4\\
1&0&1&0&1&-2&-2&4\\
1&1&0&0&1&2&-2&0\\
1&1&1&0&1&0&-2&0\\
\hline
\end{tabular}
\end{center}

\subsection{Preliminaries: The Deutsch-Jozsa Algorithm}
Given $f$ is either constant or balanced, one may ask for an algorithm,
that can answer what exactly it is. In this case the Boolean function
$f$ is available in the form of an oracle (black box), where one can
apply an input to the black box to get the output. A classical algorithm needs
to check the function for $2^{n-1} + 1$ many inputs in worst case to decide
whether the function is constant or balanced.

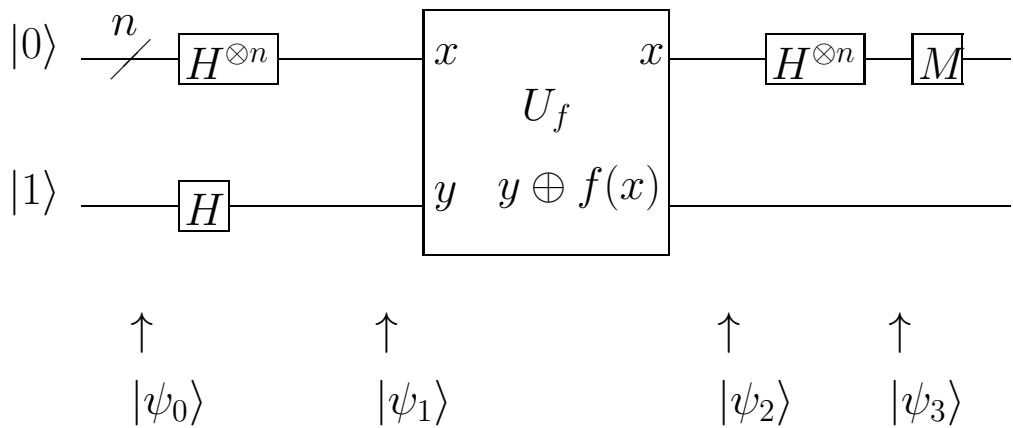
\begin{figure}[h]
\setlength{\unitlength}{6.5mm}
\begin{picture}(5,10)
\put(1.5,8.8){\makebox(0,0)[bl]{\Large {$|0\rangle$}}}
\put(1.5,5.8){\makebox(0,0)[bl]{\Large {$|1\rangle$}}}
\put(10,5){\framebox(5,5)}
\put(3,6){\line(1,0){2}}
\put(3.6,8.6){\line(1,1){0.8}}
\put(3.6,9.5){\makebox(0,0)[bl]{\Large {$n$}}}
\put(5,5.5){\framebox(1,1)}
\put(5.1,5.6){\makebox(0,0)[bl]{\Large {$H$}}}
\put(6,6){\line(1,0){4}}
\put(15,6){\line(1,0){7}}
\put(3,9){\line(1,0){2}}
\put(5,8.5){\framebox(2,1)}
\put(5.1,8.6){\makebox(0,0)[bl]{\Large {$H^{\otimes n}$}}}
\put(7,9){\line(1,0){3}}
\put(15,9){\line(1,0){2}}
\put(17,8.5){\framebox(2,1)}
\put(17.1,8.6){\makebox(0,0)[bl]{\Large {$H^{\otimes n}$}}}
\put(19,9){\line(1,0){1}}
\put(20,8.5){\framebox(1,1)}
\put(20.1,8.6){\makebox(0,0)[bl]{\Large {$M$}}}
\put(21,9){\line(1,0){1}}

\put(10.2,5.9){\makebox(0,0)[bl]{\Large {$y$}}}
\put(10.2,8.9){\makebox(0,0)[bl]{\Large {$x$}}}
\put(14.4,8.9){\makebox(0,0)[bl]{\Large {$x$}}}
\put(11.5,5.9){\makebox(0,0)[bl]{\Large {$y \oplus f(x)$}}}
\put(12,7.5){\makebox(0,0)[bl]{\Large {$U_f$}}}

\put(4,3){\makebox(0,0)[bl]{{\Large $\uparrow$}}}
\put(9,3){\makebox(0,0)[bl]{{\Large $\uparrow$}}}
\put(16,3){\makebox(0,0)[bl]{{\Large $\uparrow$}}}
\put(19.5,3){\makebox(0,0)[bl]{{\Large $\uparrow$}}}
\put(4,1.5){\makebox(0,0)[bl]{{\Large $| \psi_0 \rangle$}}}
\put(9,1.5){\makebox(0,0)[bl]{{\Large $| \psi_1 \rangle$}}}
\put(16,1.5){\makebox(0,0)[bl]{{\Large $| \psi_2 \rangle$}}}
\put(19.5,1.5){\makebox(0,0)[bl]{{\Large $| \psi_3 \rangle$}}}
\end{picture}
\caption{Quantum circuit to implement Deutsch-Jozsa Algorithm}
\label{fig1}
\end{figure}

Now we discuss the quantum computational model.
It is known that given a classical circuit $f$, there is a
quantum circuit of comparable efficiency which computes the transformation
$U_f$ that takes input like $|x, y\rangle$ and produces output like
$|x, y \oplus f(x)\rangle$. Given such an $U_f$ is available,
Deutsch-Jozsa~\cite{qDJ92} provided a quantum algorithm that can solve this
problem in constant time. We first present how the quantum circuit looks
like in Figure~\ref{fig1} and then explain the algorithm in 
Algorithm~\ref{algo1}.

\begin{algorithm} {\em Deutsch-Jozsa Algorithm~\cite{qDJ92}}
\label{algo1}
\begin{center}
\begin{tabular}{|ll|}
\hline
1. & $|\psi_0\rangle = |0\rangle^{\otimes n} |1\rangle$\\
2. & $|\psi_1\rangle = \sum_{x \in \{0, 1\}^n}
\frac{|x\rangle}{\sqrt{2^n}}\left[ \frac{|0\rangle - |1\rangle}
{\sqrt{2}}\right]$\\
3. & $|\psi_2\rangle = \sum_{x \in \{0, 1\}^n}
\frac{(-1)^{f(x)}|x\rangle}{\sqrt{2^n}}\left[ \frac{|0\rangle - |1\rangle}
{\sqrt{2}}\right]$\\
4. & $|\psi_3\rangle = \sum_{z \in \{0, 1\}^n}\sum_{x \in \{0, 1\}^n}
\frac{(-1)^{x \cdot z \oplus f(x)}|z\rangle}{2^n}\left[ \frac{|0\rangle - |1\rangle}
{\sqrt{2}}\right]$\\
5. & {\rm Measurement at $M$: all zero state implies that the function 
		is constant,}\\
\  & ~~~~~~~~~~~~~{\rm otherwise it is balanced.}\\
\hline
\end{tabular}
\end{center}
\end{algorithm}

In the next section we will keep the Algorithm~\ref{algo1} as it is 
and interpret the Step 5 of it according to our need.

\section{Problems in EQP}
\label{corr}
Let us start with some technical results on hardness of calculating the 
Walsh spectra.
\begin{proposition}
\label{proph}
A Boolean function $f$ is available in the form of an oracle. 
\begin{enumerate}
\item SAT is Turing reducible to computing Walsh transform 
at the point $0$. 
\item Finding $W_f(0)$ is outside ${\mbox P}^{\mbox{NP}}$. 
\item Given a non zero $\omega$, finding $W_f(\omega)$ is outside 
${\mbox P}^{\mbox{NP}}$. 
\end{enumerate}
\end{proposition}
\pf The function $f$ is not satisfiable, iff $W_f(0) = 2^n$. This proves
item 1. 

Now we prove item 2. In~\cite{GP01}, the following problem has been 
presented which is outside ${\mbox P}^{\mbox{NP}}$. 
A Boolean function $f$ with $wt(f)$ either $2^{n-2}$ or $3\cdot 2^{n-2}$
is given in the form of an oracle. One has to identify which one is this.
Note that $wt(f) = 2^{n-2}$ iff $W_f(0) = 2^{n-1}$ and 
$wt(f) = 3 \cdot 2^{n-2}$ iff $W_f(0) = -2^{n-1}$. Thus the result. 

The proof of item 3 is as follows. $W_f(0) = 
W_{f \oplus \omega \cdot x}(\omega)$. If the oracle of $f$ is available,
then it is easy to construct the oracle of $f \oplus \omega \cdot x$.
Hence the proof. \qed

We have already discussed in Algorithm~\ref{algo0} that the best known
classical algorithm for calculating the Walsh spectra of an $n$-variable
Boolean function requires the truth table of size $2^n$ as an input 
and then the algorithm requires $O(2^n)$ time.
Let us now describe our interpretation of Deutsch-Jozsa Algorithm 
in terms of Walsh spectra.
Note that $\sum_{z \in \{0, 1\}^n}\sum_{x \in \{0, 1\}^n}
\frac{(-1)^{x \cdot z \oplus f(x)}|z\rangle}{2^n}
= \sum_{z \in \{0, 1\}^n} \frac{W_f(z)}{2^n} |z\rangle$, i.e., the associated 
probability with a state $|z\rangle$ is $\frac{W_f^2(z)}{2^{2n}}$. Hence we
have the following result.
\begin{proposition}
\label{prop1}
Given an $n$-variable Boolean function $f$,
the Deutsch-Jozsa algorithm (Algorithm~\ref{algo1}) produces a super position 
of all the states $z \in \{0, 1\}^n$ at the measurement point $M$ with 
amplitude $\frac{W_f(z)}{2^n}$ corresponding to each state $z$.
\end{proposition}

Now let us describe the 
following problem which has been presented in~\cite{BV93} as parity problem.

\begin{problem}{\em ~\cite{BV93}}
Let $f$ be an linear $n$-variable Boolean function, i.e., 
$f(x) = \omega \cdot x$, available in the form of an oracle, 
Find out the $\omega$. 
\end{problem}

For a linear function $f(x) = \omega \cdot x$, 
$W_f(\omega) = 2^n$ and $W_f(z) = 0$, for $z \neq \omega$. Thus the observed 
state of $n$ bits in the Step 5 of Algorithm~\ref{algo1} will clearly
output $\omega$ itself (with probability $\frac{W_f^2(\omega)}{2^{2n}} = 1$).
Thus the Deutsch-Jozsa algorithm solves this problem in constant time.
In classical model this problem clearly needs $O(n)$ time.
This difference has been exploited and it has been shown that BPP is not equal 
to BQP with respect to an oracle~\cite{BV93}.

Now we present the problem we described.
\begin{problem}
\label{prob2}
A Boolean function $f$ is given in the form of an oracle. 
Find out an $\omega$, such that $W_f(\omega) \neq 0$.
\end{problem}
The solution to this problem using the Deutsch-Jozsa algorithm works as 
follows. Let us consider that $S = \{\omega | W_f(\omega) \neq 0\}$. For
any $\omega \in \{0, 1\}^n \setminus S$, $W_f(\omega) = 0$. 
Note that for $\omega \in S$,
$\sum_{x \in \{0, 1\}^n} (-1)^{f(x) \oplus x \cdot \omega}$ is nonzero and
for $\omega \in \{0, 1\}^n \setminus S$,
$\sum_{x \in \{0, 1\}^n} (-1)^{f(x) \oplus x \cdot \omega}$ is zero.
We have already discussed that the associated probability
with a state $|z\rangle$ is $\frac{W_f^2(z)}{2^{2n}}$.
Here the probability associated with $|z\rangle$ is nonzero when $z \in S$ and 
the probability associated with $|z\rangle$ is $0$ when 
$z \in \{0, 1\}^n \setminus S$. It is clear that the sum of probabilities 
associated with the states in $S$ is $1$. Thus, the state, say $\omega$, 
observed after the measurement at Step 5 belongs to $S$ and for the 
observed $\omega$, $W_f(\omega) \neq 0$. Hence the Problem~\ref{prob2} can be 
solved in constant time using the Deutsch-Jozsa algorithm.

Based on the above discussion we have the following result.
\begin{theorem}
\label{th1}
The Problem~\ref{prob2} belongs to EQP with respect to the oracle $f$.
\end{theorem}

Now we present a related problem where one needs to find out the maximally 
correlated linear or affine function with respect to $f$.
\begin{problem}
\label{prob3}
A Boolean function $f$ is given in the form of an oracle.
Find out an ${\hat \omega}$, such that $|W_f({\hat \omega})| 
= \max_{\omega \in \{0, 1\}^n}|W_f(\omega)|$.
\end{problem}
Algorithm~\ref{algo1} does not guarantee the answer to 
Problem~\ref{prob3}. Since Algorithm~\ref{algo1} is probabilistic in nature,
it may very well happen that it outputs some $\omega^\prime$, for which 
$W_f(\omega^\prime) \neq 0$, but $|W_f(\omega^\prime)| < |W_f({\hat \omega})|$. 
That means we get a linear or affine function which is correlated to
$f$, but not maximally correlated. 

There exists a sub class of Boolean functions, the bent 
functions~\cite{RO76}, 
for which one can solve Problems~\ref{prob2},~\ref{prob3} in one step 
using classical computational model also. For a bent function $f$, 
$W_f(\omega) = \pm 2^{\frac{n}{2}}$, for any $\omega \in \{0, 1\}^n$.
Thus if it is known that the function is a bent function, then one can 
choose any $\omega$ and produce that as the output. However, it is very clear
these problems are not easy in general.

One very interesting class of Boolean functions are the ones
where the Walsh spectra become three valued $0, \pm 2^k$. 
These functions are referred as plateaued functions
in literature~\cite{ZZ99,CC03}. The class of plateaued functions contains
cryptographically significant Boolean functions, including certain classes
of resilient functions~\cite{ZZ99,SM00B,CC03} and hence these functions are
actually used in crypto systems. Now consider the following problem.
\begin{problem}
\label{prob4}
A plateaued Boolean function $f$ (i.e., $W_f(\omega)$ can take the values
$0, \pm 2^k$) is given in the form of an oracle.
Find out an ${\hat \omega}$, such that $|W_f({\hat \omega})| 
= \max_{\omega \in \{0, 1\}^n}|W_f(\omega)|$, which is equivalent to
find out an ${\hat \omega}$, such that $W_f({\hat \omega}) \neq 0$.
\end{problem}
Clearly Algorithm~\ref{algo1} outputs proper solution in one step, but the
best known classical algorithm till date which can deterministically
solve this problem is the Fast Walsh transform which requires $O(n2^n)$ time
in worst case. The information that the Walsh spectra is three valued
does not help in the calculation of Walsh spectra in a better way on the 
classical model.

There are different kinds of resilient, correlation immune 
and other cryptographically significant Boolean 
functions~\cite{ZZ99,SM00A,SM00B,CC02,CC03} with three valued Walsh spectra.
These functions are used for robust design of crypto systems. 
Getting a linear or affine function which is maximally correlated to 
the Boolean function in constant time directly helps in cryptanalysis of 
such crypto systems and presents an application to Algorithm~\ref{algo1},
the Deutsch-Jozsa Algorithm~\cite{qDJ92}.

We further restrict the Problem~\ref{prob4} and present the following problem
to highlight the exponential speed up of quantum algorithms over classical
domain.
\begin{problem}
\label{prob5}
A plateaued $n$-variable ($n$ odd) Boolean function $f$ with three 
valued Walsh spectra $0, \pm 2^{\frac{n+1}{2}}$ is given in the form of an 
oracle. Find out an ${\omega}$, such that $W_f({\omega}) \neq 0$.
\end{problem}
Algorithm~\ref{algo1} solves this problem in one step. 
\begin{theorem}
\label{th2}
Problem~\ref{prob5} belongs to EQP with respect to the oracle $f$.
\end{theorem}
The best known classical algorithm, fast Walsh transform, needs $O(n2^n)$ 
time and the structure of the problem does not reveal anything to present 
a better deterministic classical algorithm. To analyse the situation in more 
details, let us define {\em restricted Walsh transform}.
The {\em restricted Walsh transform} of $f({x})$ on a subset $T$ of 
$\{0, 1\}^n$ is a real valued function
over $\{0, 1\}^n$ which is defined as $$W_f({\omega})|_T = \sum_{{x} 
\in T} (-1)^{f({x}) \oplus {x} \cdot {\omega}}.$$
Any NP machine can guess an $\omega$ but it is impossible to verify in 
polynomial time whether the value of Walsh spectra at chosen $\omega$ is non 
zero. This is because $f$ is presented as a black box and thus one needs to 
query the value of $f$ in at least $2^{n-1} + 1$ times at the best case
to decide whether $W_f(\omega)$ is non zero. 
Let $T \subset \{0, 1\}^n$ such that $|T| = 2^{n-1} + 1$.
If one finds that $W_f(\omega)|_T$ is $2^{n-1} + 1$ or $-2^{n-1} - 1$,
then it is clear that $W_f(\omega)$ cannot be zero. However, it is not possible
to decide whether $W_f(\omega)$ is $0$ or $\pm 2^{\frac{n+1}{2}}$ 
from $W_f(\omega)|_T$ when $|T| \leq 2^{n-1}$. Thus the verification
stage needs $O(2^n)$ many queries to the oracle at the best case. 

Though we can not present any formal proof, it seems that Problem~\ref{prob5}
is outside BPP (may be even outside PH) with respect to the oracle $f$ 
and once such a result can be proved, the Deutsch-Jozsa algorithm can be
used to present a relativized separation between BPP (may be PH) and EQP.
This we place as an important open problem in this direction.

\section{Problems in BQP}
\label{bqp}
Let us consider a subset of Boolean functions with the following property.
$${\cal L}_n = \{f \in \Omega_n | d(f, l) \leq 2^{n-3}, l \in L(n)\}.$$

\begin{proposition}
$|{\cal L}_n| = 2^n \sum_{i = 0}^{2^{n-3}} \comb{2^n}{i}$.
\end{proposition}
\pf Let ${\cal L}_n^l = \{f \in \Omega_n | d(f, l) \leq 2^{n-3}\}$. Since
for distinct $l_1, l_2 \in L(n)$, $d(l_1, l_2) = 2^{n-1}$, we have
${\cal L}_n^{l_1} \cap {\cal L}_n^{l_2} = \emptyset$. Also it is clear that
$|{\cal L}_n^{l_1}| = |{\cal L}_n^{l_2}|$. Since, $|L(n)| = 2^n$, 
and ${\cal L}_n = \cup_{l \in L(n)} {\cal L}_n^l$,
$|{\cal L}_n| = 2^n |{\cal L}_n^l|$ for some $l \in L(n)$. Now 
$|{\cal L}_n^l| = \sum_{i = 0}^{2^{n-3}} \comb{2^n}{i}$ as one can choose
$i$ $(0 \leq i \leq 2^{n-3})$ many positions in the truth table of the 
linear function $l$ and complement them to get an $f$. This gives the
proof. \qed

From~\cite[Page 165]{HM80}, $\sum_{i = 0}^{\lambda u} \binom{u}{i} 
\leq 2^{u H(\lambda)},$ where the binary entropy function
$H(\lambda) = -\lambda \log_2 \lambda
- (1 - \lambda) \log_2 (1 - \lambda).$ Also it is clear that
$\sum_{i = 0}^{2^{n-3}} \comb{2^{n-3}}{i} < 
\sum_{i = 0}^{2^{n-3}} \comb{2^n}{i}$. Thus, 
$2^{2^{n-3}} < |{\cal L}|_n =  \sum_{i = 0}^{2^{n-3}} \comb{2^n}{i} 
\leq 2^{2^n H(\frac{1}{8})}$.

Let us consider the following problem which is a restricted version of
Problem~\ref{prob3}.
\begin{problem}
\label{prob6}
An $n$-variable ($n$ odd) Boolean function $f \in {\cal L}_n$ 
is given in the form of an oracle. 
Find out an ${\hat \omega}$, such that $|W_f({\hat \omega})| 
= \max_{\omega \in \{0, 1\}^n}|W_f(\omega)|$.
\end{problem}

\begin{lemma}
Problem~\ref{prob6} belongs to BQP with respect to the oracle $f$.
\end{lemma}
\pf If $f \in {\cal L}_n$, then $d(f, {\hat \omega}\cdot x) \leq 2^{n-3}$, 
i.e., $W_f({\hat \omega}) \geq 
2^n - 2 d(f, {\hat \omega}\cdot x) = 2^{n} - 2^{n-2}$.
Thus the success probability of Algorithm~\ref{algo1} is 
$\geq (\frac{2^n-2^{n-2}}{2^n})^2 = \frac{9}{16}$. 
The probability of getting a wrong answer is 
$\leq \frac{7}{16}$. \qed

Now we refine these results a little bit to extend the class ${\cal L}_n$. 
Let $${\cal L}_{n, \epsilon} = 
\{f \in \Omega_n | d(f, l) \leq (1+(3-2\sqrt{2}-4\epsilon)) 
2^{n-3}, l \in L(n), 0 < \epsilon < \frac{3-2\sqrt{2}}{4}\}.$$
It is clear that 
$|{\cal L}_{n, \epsilon}| > |{\cal L}_{n}|$ as
$3-2\sqrt{2}-4\epsilon > 0$, for the given range of $\epsilon$.

If $f \in {\cal L}_{n, \epsilon}$, then 
$d(f, {\hat \omega}\cdot x) \leq (1+(3-2\sqrt{2}-4\epsilon)) 2^{n-3}$, 
i.e., $W_f({\hat \omega}) \geq 
2^n - 2 d(f, {\hat \omega}\cdot x) = 2^{n} - 2^{n-2}(4-2\sqrt{2}-4\epsilon)$.
Thus the success probability of Algorithm~\ref{algo1} is 
$\geq (\frac{2^{n-2}(2\sqrt{2}+4\epsilon)}{2^n})^2 = 
(\frac{1}{\sqrt{2}}+\epsilon)^2
= \frac{1}{2} + \sqrt{2}\epsilon + \epsilon^2 > \frac{1}{2} + \epsilon$. 
The probability of getting a wrong answer is 
$< \frac{1}{2} - \epsilon$. 

Noting $\sqrt{2} < 1.415$, one can use a small constant 
$\epsilon$ such that
$${\cal L}_{n, \epsilon} = 
\{f \in \Omega_n | d(f, l) \leq 1.17 \cdot 2^{n-3}, l \in L(n)\}.$$

Based on the above results we present the following problem and corollary.

\begin{problem}
\label{prob7}
An $n$-variable ($n$ odd) Boolean function $f \in {\cal L}_{n, \epsilon}$ 
is given in the form of an oracle. 
Find out an ${\hat \omega}$, such that $|W_f({\hat \omega})| 
= \max_{\omega \in \{0, 1\}^n}|W_f(\omega)|$.
\end{problem}

\begin{corollary}
Problem~\ref{prob7} belongs to BQP with respect to the oracle $f$.
\end{corollary}

To the best of our knowledge, there is no other way to solve 
Problem~\ref{prob6} and Problem~\ref{prob7} deterministically in classical 
domain without calculating the Walsh spectra. 

\section{Conclusion}
In this note, we identify a large set of problems which are in EQP or BQP
with respect to an oracle $f$, where $f$ is an $n$-variable Boolean function
available in the form of a black box. We have used the basic 
Deutsch-Jozsa algorithm to prove our results and show further applications
to this well known algorithm. The only known tool to solve these problems
in classical computational model is calculation of Walsh spectra which requires
$O(n2^n)$ time. It is left open whether these problems are indeed hard to 
solve from complexity theoretic viewpoint. If that can be shown then the
problems mentioned here, along with the Deutsch-Jozsa algorithm can be used to 
prove important results related to relativized separation between 
BPP (may be PH) and EQP or BQP.


\begin{thebibliography}{10}

\bibitem{SA02}
S.~Aaronson.
\newblock Quantum Lower Bound for Recursive Fourier Sampling.
\newblock Available at http://www.lanl.gov/list/quant-ph/0209060,
9 September, 2002.

\bibitem{BV93}
E.~Bernstein and U.~Vazirani.
\newblock Quantum complexity theory.
\newblock In Proceedings of 25th Annual ACM Symposium on Theory of Computing,
1993, pages 11--20.

\bibitem{BV97}
E.~Bernstein and U.~Vazirani.
\newblock Quantum complexity theory.
\newblock {\em SIAM Journal of Computing}, 26(5):1411--1473, 1997.

\bibitem{AM00}
A.~Canteaut and M.~Trabbia.
\newblock Improved fast correlation attacks using parity-check equations of
 weight 4 and 5.
\newblock In {\em Advances in Cryptology - EUROCRYPT 2000}, number 1807 in
  Lecture Notes in Computer Science, pages 573--588. Springer Verlag, 2000.

\bibitem{CC02}
C.~Carlet.
\newblock 
A larger Class of Cryptographic Boolean Functions via a Study of the
Maiorana-McFarland Constructions.
\newblock In {\em Advances in Cryptology - CRYPTO 2002}, number 2442 in Lecture
  Notes in Computer Science, pages 549--564. Springer Verlag, 2002.

\bibitem{CC03}
C.~Carlet and E.~Prouff.
\newblock On plateaued functions and their constructions.
\newblock In {\em FSE 2003}, number 2887 in Lecture
  Notes in Computer Science, pages 54--73. Springer Verlag, 2003.

\bibitem{qDJ92}
D.~Deutsch and R.~Jozsa.
\newblock Rapid solution of problems by quantum compuation.
\newblock Proceedings of Royal Society of London, A439:553--558 (1992).

\bibitem{CD91}
C.~Ding, G.~Xiao, and W.~Shan.
\newblock {\em The Stability Theory of Stream Ciphers (Book)}.
\newblock Number 561 in Lecture Notes in Computer Science. Springer-Verlag,
  1991.

\bibitem{DK00}
D.~-Z.~Du and K.~-I.~Ko.
\newblock {\em Theory of Computational Complexity (Book)}.
\newblock John Wiley \& Sons, INC., 2000.

\bibitem{GP01}
F.~Green and R.~Pruim.
\newblock Relativized Separation of EQP from ${\mbox P}^{\mbox{NP}}$. 
\newblock {\em Information Processing Letters}, 80(5):257--260, 2001.

\bibitem{qGR96}
L.\ Grover.
\newblock A fast quantum mechanical algorithm for database search.
\newblock In {\em Proceedings of 28th Annual Symposium on the Theory of 
Computing (STOC)}, May 1996, pages 212--219. Available at 
xxx.lanl.gov/quant-ph/9605043.

\bibitem{XZ88}
X.~Guo-Zhen and J.~Massey.
\newblock A spectral characterization of correlation immune combining
  functions.
\newblock {\em IEEE Transactions on Information Theory}, 34(3):569--571, May
  1988.

\bibitem{HM80}
R.~W.~Hamming.
\newblock {\em Coding and Information Theory}.
\newblock Prentice-Hall, Inc., Englewood Cliffs, N. J. 07632, 1980.

\bibitem{TH00B}
T.~Johansson and F.~Jonsson.
\newblock Fast correlation attacks through reconstruction of linear
  polynomials.
\newblock In {\em Advances in Cryptology - CRYPTO 2000}, number 1880 in Lecture
  Notes in Computer Science, pages 300--315. Springer Verlag, 2000.

\bibitem{MM94}
M.~Matsui.
\newblock Cryptanalysis method for DES cipher.
\newblock In {\em Advances in Cryptology, Eurocrypt 1993}, Lecture Notes in
Computer Science, Number 765, Pages 386--397, Springer-Verlag, 1994.

\bibitem{JJ83}
J.~J. Mykkeltveit.
\newblock The covering radius of the $(128, 8)$ {R}eed-{M}uller code is 56.
\newblock {\em IEEE Transactions on Information Theory}, IT-26(3):358--362,
  1983.

\bibitem{qNC02}
M.~A.~Nielsen and I.~L.~Chuang.
\newblock Quantum Computation and Quantum Information.
\newblock Cambridge University Press, 2002.

\bibitem{PW83}
N.~J. Patterson and D.~H. Wiedemann.
\newblock The covering radius of the $(2^{15}, 16)$ {R}eed-{M}uller code is at
  least 16276.
\newblock {\em IEEE Transactions on Information Theory}, IT-29(3):354--356,
  1983. See correction at IT-36(2):443, 1990.

\bibitem{SM00E}
E.~Pasalic, S.~Maitra, T.~Johansson and P.~Sarkar.
\newblock New constructions of resilient and correlation immune {B}oolean
  functions achieving upper bounds on nonlinearity.
\newblock In {\em Workshop on Coding and Cryptography - WCC 2001}, Paris, 
January 8--12, 2001. Electronic Notes in Discrete Mathematics, Volume 6,
Elsevier Science, 2001.

\bibitem{RO76}
O.~S. Rothaus.
\newblock On bent functions.
\newblock {\em Journal of Combinatorial Theory, Series A}, 20:300--305, 1976.

\bibitem{SM00A}
P.~Sarkar and S.~Maitra.
\newblock Construction of nonlinear {B}oolean functions with important
  cryptographic properties.
\newblock In {\em Advances in Cryptology - EUROCRYPT 2000}, number 1807 in
  Lecture Notes in Computer Science, pages 485--506. Springer Verlag, 2000.

\bibitem{SM00B}
P.~Sarkar and S.~Maitra.
\newblock Nonlinearity bounds and constructions of resilient Boolean functions.
\newblock In {\em Advances in Cryptology - CRYPTO 2000}, number 1880 in Lecture
  Notes in Computer Science, pages 515--532. Springer Verlag, 2000.

\bibitem{SG84}
T.~Siegenthaler.
\newblock Correlation-immunity of nonlinear combining functions for
  cryptographic applications.
\newblock {\em IEEE Transactions on Information Theory}, IT-30(5):776--780,
  September 1984.

\bibitem{SG85}
T.~Siegenthaler.
\newblock Decrypting a class of stream ciphers using ciphertext only.
\newblock {\em IEEE Transactions on Computers}, C-34(1):81--85, January 1985.

\bibitem{ZZ99}
Y.~Zheng and X.~M. Zhang.
\newblock Plateaued functions.
\newblock In {\em ICICS'99}, number 1726 in Lecture Notes in Computer Science,
  pages 284--300. Springer Verlag, 1999.

\end{thebibliography}
\end{document}